\documentclass{article}
\usepackage{cite} 
\usepackage{ijcai25}
\usepackage{times}
\usepackage{soul}
\usepackage{url}
\usepackage[hidelinks]{hyperref}
\usepackage[utf8]{inputenc}
\usepackage[small]{caption}
\usepackage{graphicx}
\usepackage{amsmath}
\usepackage{diagbox}
\usepackage{amsthm}
\usepackage{booktabs}
\usepackage{algorithm}
\usepackage{algorithmic}
\usepackage[switch]{lineno}
\usepackage{color}
\usepackage{amsfonts}

\title{Lend a Hand: Semi Training-Free Cued Speech Recognition via MLLM-Driven Hand Modeling for Barrier-free Communication}

\author{
Guanjie Huang$^1$
\and
Danny Hin Kwok Tsang$^{1,2}$\And
Li Liu$^{1,2,}$\thanks{Corresponding author.}
\affiliations
$^1$ The Hong Kong University of Science and Technology (Guangzhou)
\\
$^2$ The Hong Kong University of Science and Technology\\
}

\begin{document}

\maketitle
\begin{abstract}
Cued Speech (CS) is an innovative visual communication system that integrates lip-reading with hand coding, designed to enhance effective communication for individuals with hearing impairments. Automatic CS Recognition (ACSR) refers to the AI-driven process of automatically recognizing hand gestures and lip movements in CS, converting them into text. However, previous work often relies on complex fusion modules and training techniques. Additionally, due to the limited amount of data in CS, the extraction of hand features, as well as recognition modeling, has consistently been subpar, significantly limiting the effectiveness of ACSR. To address this issue, we have innovatively explored the capabilities of Multimodal large language models (MLLMs) in recognizing hand shapes and positions in CS. More precisely, we propose a new \textbf{S}emi \textbf{T}raining-\textbf{F}ree paradigm for \textbf{ACSR}, named \textbf{STF-ACSR}. This approach leverages zero-shot recognition of hand movements through the \textbf{C}hinese \textbf{CS} \textbf{P}rompt \textbf{M}odule \textbf{(CCSPM)} which equipped a training-free keyframe filtering and customized prompt engineering based on MLLM. It then integrates the recognition results into the lip-reading model using a \textbf{Minimalist Fusion Module (MFM)}, effectively achieving superior recognition results. Furthermore, specifically for this study, 
we have supplemented the existing dataset of 6 normal hearing CS cuers\footnote{People who perform CS are called cuers.} by recording additional data from 8 cuers with hearing impairments, resulting in a new mixed dataset. Extensive experiments have demonstrated that STF-ACSR significantly outperforms previous methods on both normal and hearing-impaired data. Implementation and checkpoints are available at \href{https://github.com/DennisHgj/STF_ACSR}{\textit{https://github.com/DennisHgj/STF\_ACSR}}.
\end{abstract}

\section{Introduction}
\label{sec:intro}

Hearing-impaired people often struggle with lip-reading due to limited information. In 1967, \cite{cued1967} invented the first Cued Speech (CS) system for American English. It uses hand codings to help with lip-reading at the phonetic level, making spoken language visible.
\begin{figure}[t]
\includegraphics[width=1.0\linewidth]{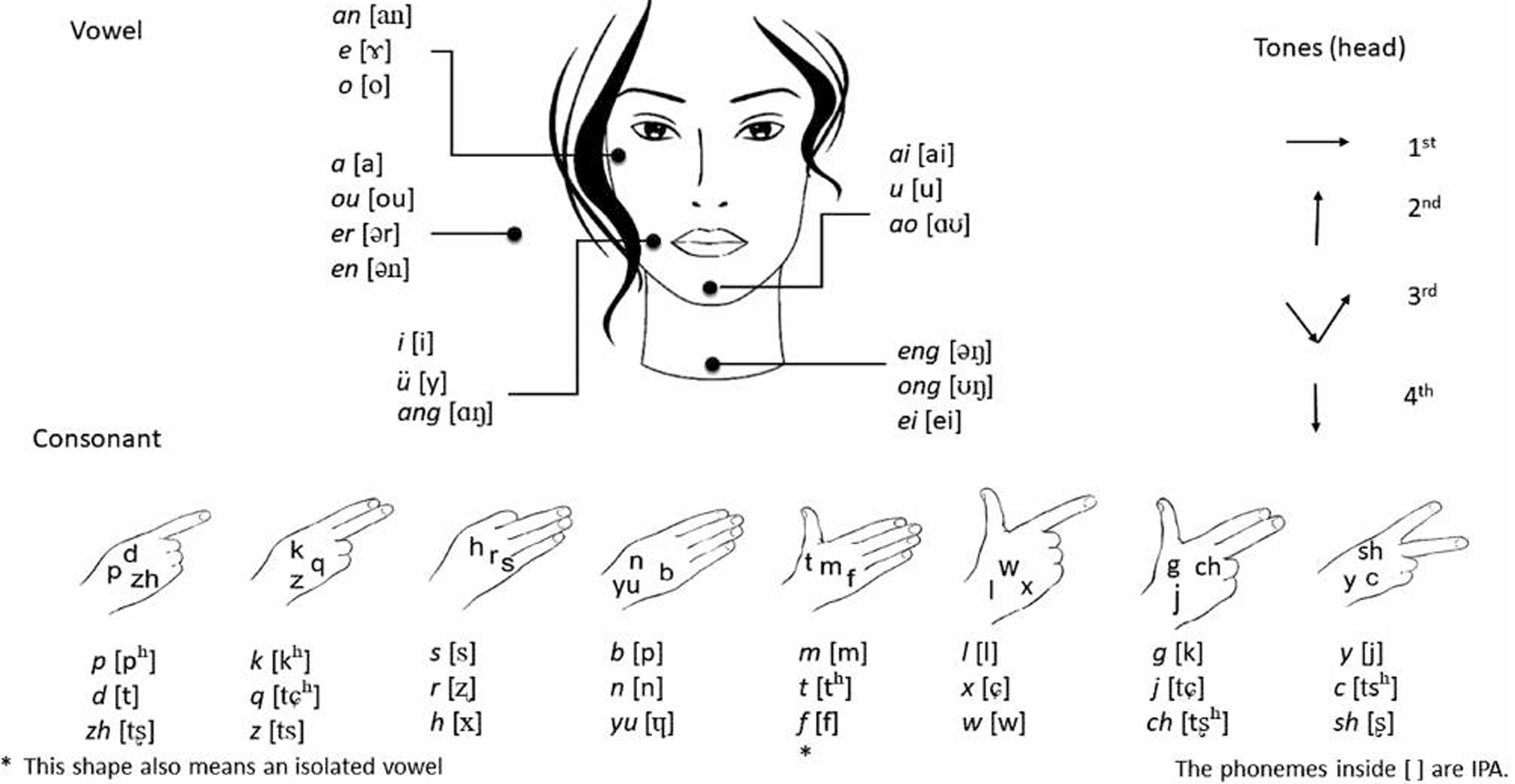}
\centering
\caption{\textbf{Mandarin Chinese CS system}. Five hand positions (mouth, chin, throat, side, cheek) and eight hand shapes are defined to encode Chinese vowels and consonants to assist lip reading (image from \protect\cite{liu2019pilot}).}
\label{fig:mandarinCS} 
\end{figure}

Since its invention, CS has demonstrated remarkable adaptability and has been successfully adapted to over 65 spoken languages. 
In 2019, \cite{liu2019pilot} presented the first Mandarin Chinese CS system (shown in Fig.~\ref{fig:mandarinCS}). In this system, a set of five hand positions is defined to represent all Chinese vowel groups. Complementing this, eight hand shapes are defined to encode 24 Chinese consonant phonemes. This development has opened up new possibilities for hearing-impaired Mandarin speakers, enabling them to access better and understand the spoken language through visual cues.

\textcolor{black}{As deep learning (DL) emerged, Automatic CS Recognition (ACSR)~\cite{liu2018automatic} has drawn more interest, due to its potential to offer great help to the hearing-impaired in their day-to-day communication. The objective of ACSR is to transform multi-modal inputs of CS videos (i.e., lip and hand) into text. For this task, the current trend based on transformers, tries to form a suitable cross-modal fusion strategy~\cite{liu2023cross,liu2024computation} to deal with the complementary relationships that come from these multi-modal inputs, ensuring the effective conversion from input to text output. However, the limited scale of ACSR datasets hinders the training of complex fusion modules, leading to suboptimal performance and significant performance degradation in hearing-impaired scenarios.}

\begin{figure*}[t]
\includegraphics[width=1.0\linewidth]{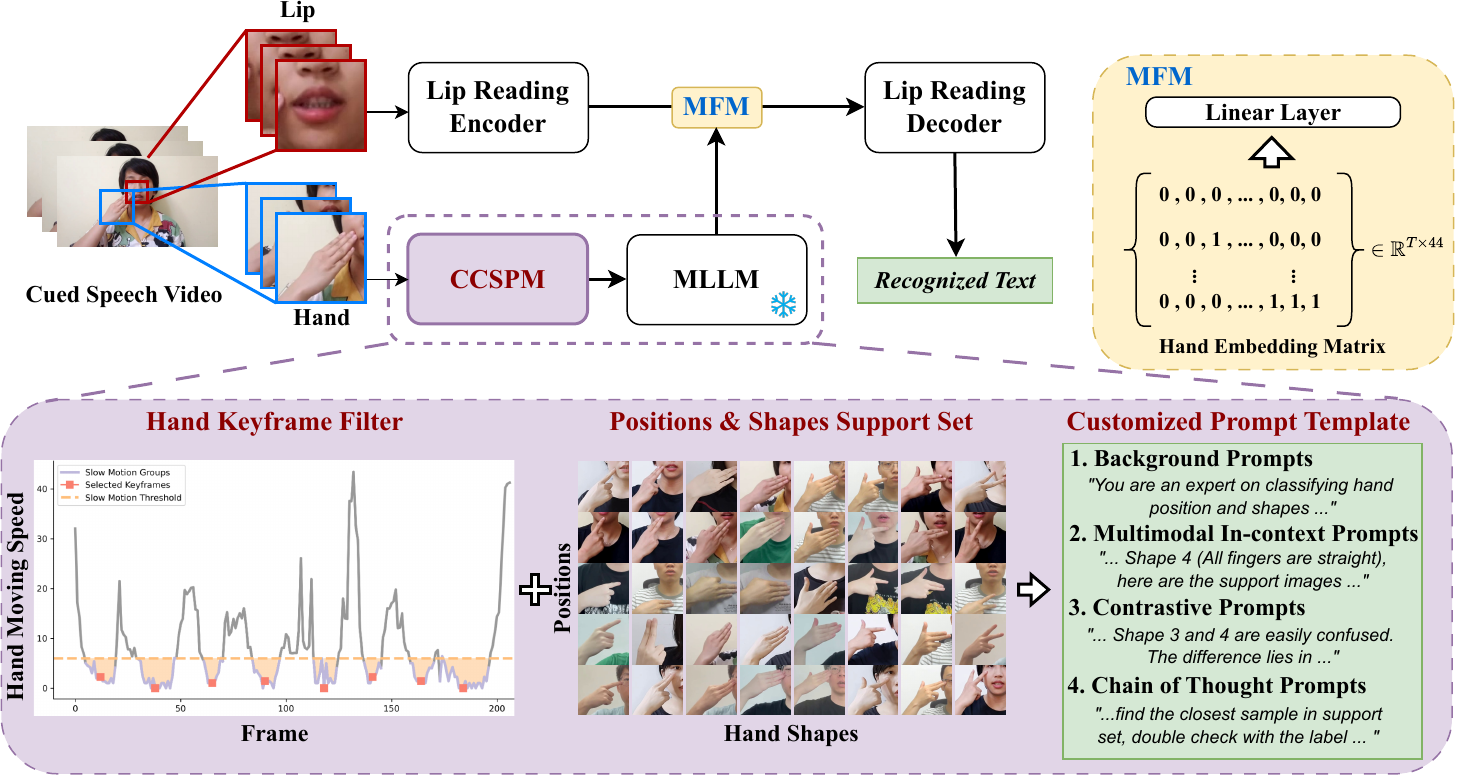}
\centering
\caption{\textbf{Overview of STF-ACSR.}  It extracts hand information with the help of CCSPM and fuses it with lip features through MFM. CCSPM consists of a Hand Keyframe Filter, a positions and shapes support set, and a customized prompt template. MFM embeds the recognition results of MLLM and adjusts the dimension through a linear layer to achieve the fusion of hand and lip information.  ``CCSPM'',  ``MFM'', and ``MLLM'' are shorts for the Chinese Cued Speech Prompt Module, Minimalist Fusion Module, and Multimodal Large Language Model. For the hand embedding matrix, $T$ indicates the frame number, 44 is composed of 40 phonemes and four symbols.
}
\label{fig:framework} 
\end{figure*}  
To achieve stable performance improvement in different scenarios without relying on complex fusion modules, in this work, 
we propose a new \textbf{S}emi \textbf{T}raining-\textbf{F}ree paradigm for \textbf{ACSR}, named \textbf{STF-ACSR} (see Fig.~\ref{fig:framework}).
More precisely, to avoid the overfitting of the hand recognition, we propose a Chinese CS Prompt Module (CCSPM), which leverages and motivates the ability of Multimodal Large Language Model (MLLM) to recognize the CS hand shape and position in a training-free manner. 



The CCSPM first presents a hand keyframe filter based on the Chinese CS system, guided by two main rules. The first is the word rule, which indicates that a Chinese word's phoneme combination consists of zero to two consonants and one vowel, translating to hand shapes with a consistent position by hand coding. The second rule reflects cuers' habits, where hands remain in the vowel area while maintaining the corresponding shape during communication. This filter identifies keyframes in the input video by analyzing hand movement speed to meet the token length requirements of the MLLM. Secondly, CCSPM includes a support set of 40 hand images covering all 5 positions and 8 shapes in hand coding, and a specialized prompt template to improve the accuracy of predicting hand position and shape. This template outlines the ACSR context and hand recognition goals, providing detailed descriptions of hand position and shape labels, as well as a reasoning mechanism that utilizes the support set for effective in-context reasoning.

Based on CCSPM, to reduce the complex design of the lip and hand fusion, we have mimicked the original design concept of cueing by utilizing the hand information generated by the CCSPM as a prompt for lip reading. To achieve lip and hand fusion in ACSR, we designed a simple yet effective information integration method called the Minimalist Fusion Module (MFM). It embeds the recognition results into a matrix and fuses it with lip reading features via a single linear layer.

In summary, the key contributions can be summarized as follows:

\begin{itemize}
    \item We propose a new semi training-free paradigm for ACSR (named STF-ACSR). 
   It introduces the MLLM to automatically extract hand shape and position information from CS video without needing any hand data for training. This approach moves away from the complexity of previous methods that relied on small CS datasets.
    \item Instead of using all frames, the proposed CCSPM simplifies CS hand recognition by classifying hand positions and shapes only based on keyframes. It contains a hand keyframe filter and prior-based customized prompts to leverage the visual capabilities of the MLLM, resulting in accurate CS hand understanding.
    \item We additionally record data from 8 cuers with hearing impairments, establishing the first Chinese CS dataset in the field that includes multiple individuals with hearing disabilities. Extensive experiments on the Chinese CS dataset demonstrate that the STF-ACSR significantly outperforms SOTA methods in various settings.
\end{itemize}

\section{Related Works}
\label{sec:relatedwork}
\subsection{Automatic Cued Speech Recognition} 

In 2018, \cite{liu2018automatic}~proposed this concept for the first time and used DL-based models to realize ACSR. Traditional idea believes designing a suitable fusion module can handle the natural asynchrony issue between lip and hand modals in ACSR tasks. Many works follow this idea to build ACSR systems by designing specific multimodal fusion modules and achieving good recognition results. For example, early works~\cite{heracleous2012continuous,liu2018visual,wang2021attention} extract features from the region of interest (ROI) of the lip and hand and directly concatenate them to achieve cross-modal fusion. \cite{gao2023novel} proposed a re-synchronization process for multimodal alignment, which demands the statistical pre-definition of the hand-preceding time in the CS dataset. \cite{wang2021cross} utilized knowledge distillation to draw out effective features from the teacher's knowledge of speech data. 

Recently, transformers have been widely used in processing sequential data due to their ability on capturing global information. Cross-attention that emerged to achieve multimodal fusion has also been applied to ACSR tasks. \cite{liu2023cross} proposed a hand-lip mutual learning framework based on transformers. It handles the asynchronous issue in ACSR, becoming the first work solely reliant on sentence-level annotation instead of phoneme-level annotation used in previous work. 
Subsequent work~\cite{liu2024computation} optimized the shortcomings of the transformer structure with high trainable parameters and computational complexity, and achieved better performance with a carefully designed fusion module. The above works use features extracted by pre-trained CNN front-end models to train the specially designed transformer-based fusion module from scratch.

\subsection{Lip Reading Models}

Recall that CS is a visual communication system that integrates lip reading with
hand coding. Therefore, Lip reading is an important modality in CS. In fact, Lip reading (often called Visual Speech Recognition (VSR)) has been fully developed due to its inherent advantage of abundant data in recent years. The VSR task aims to recognize spoken words by lip movements in silent videos, which is consistent with the goal of lip reading in ACSR. The VSR models achieve promising performance by integrating the hybrid CTC/Attention objective~\cite{petridis2018audio,watanabe2017hybrid} and employing transformer architectures~\cite{chang2024conformer,ma2023auto}. In addition, researchers have tried to effectively transfer or distill information from the audio modality to further improve performance~\cite{afouras2020asr,kim2022distinguishing,zinonos2023learning}. However, due to the lack of lip reading data of hearing-impaired people and the natural differences between lip reading and normal people, it is hard to obtain satisfactory recognition results by directly using the existing VSR models.

\section{Methods}
\label{sec:methods}

\subsection{Problem Formulation}
CS dataset consists of $N$ speech video and sentence-level label sequence pairs, denoted by $F =  \{[X_i,  Y_i]\}^N_{i=1}$. After preprocessing speech videos to obtain hand and lip ROI, they transfer to the lip ROI videos, hand ROI videos, and hand position sequences. The dataset now denotes to $F =  \{[X_i^{l}, X_i^{h}, X_i^{p}, Y_i]\}^N_{i=1}$. Specifically, for a speech video with $T$ frames, $X_i^{l} \in \mathbb{R}^{T\times W^l \times H^l}$ and $X_i^{h} \in \mathbb{R}^{T\times W^h \times H^h}$
where $W^l, H^l, W^h, H^h$ are the width and height of lip and hand ROI frames.
$X_i^{p} \in \mathbb{R}^{T\times 2}$, which indicates the coordinates of the center point of the hand in the original video. The problem's goal is to train a model mapping multi-modal data streams $(X^{l}, X^{h}, X^{p})$ to corresponding label sequences $Y$.

\subsection{Overview of STF-ACSR}
STF-ACSR is a semi-training-free framework inheriting the intention of the CS system design. Hand ROI frames and lip ROI frames are fed into different processing branches. As illustrated in Fig.~\ref{fig:framework}, STF-ACSR uses CCSPM to simplify the hand movement recognition task into the position and shape classification tasks for keyframes of hand ROI videos. Then, the recognition result is fused with the lip reading encoder processed lip feature through MFM, and the final text sequence is obtained through the decoder.

\subsection{Chinese CS Prompt Module}

With the innovative discovery of the rules of the Chinese CS system, CCSPM effectively simplifies the complex hand movement recognition task into the keyframe position and shape classification tasks through three components: the hand keyframe filter, the hand position and shape support set, and the customized prompt template.

\begin{algorithm}[t!]
    \caption{Hand Keyframe Filter}
    \label{alg:CKF}
    \textbf{Input}: Hand ROI frames $X^h$, Frame number $T$, Hand position sequences $X^p$\\
    \textbf{Parameter}: Slow motion threshold $\sigma$, Index distance threshold $\theta$\\
    \textbf{Function}: Distance function $distance(\cdot,\cdot)$, Get the median frame from list $median(\cdot)$\\
    \textbf{Output}: Keyframe set $K$, Slow motion group set $G$\\
    \vspace{-1.0em}
    \begin{algorithmic}[1] 
        \STATE \COMMENT{GET SLOW MOTION FRAMES}
        \STATE Let $j=1$
        \STATE Initialize slow motion frame list $S$
        \WHILE{$j < T$}
        \STATE $ D_j = distance(X_{j-1}^{p},X_j^{p})$
        \IF {$D_j\leq \sigma$}
        \STATE $S \leftarrow S \cup {X^h_j}$
        \ENDIF
        \STATE $j=j+1$
        \ENDWHILE
        \STATE \COMMENT{GET KEY FRAMES \& SLOW MOTION GROUPS}
        \STATE Let $k=0, m=0$
        \STATE Initialize current slow-motion group list $G_k$
        \FOR{$X^h_j$ in $S$}
        \IF {$G_k =\emptyset $}
        \STATE $G_k \leftarrow G_k \cup {X^h_j}, m=j$
        \ELSIF {$j-m \leq \theta$}
        \STATE $G_k \leftarrow G_k \cup {X^h_j}, m=j $
        \ELSE 
        \STATE   $K \leftarrow K \cup {median(G_k)}$
        \STATE   $G \leftarrow G \cup {G_k} , k =k+1$ 
        \ENDIF
        \ENDFOR
        \STATE \textbf{return} $K,G$
    \end{algorithmic}
\end{algorithm}

\noindent \textbf{Hand Keyframe Filter.} The overall process of Hand Keyframe Filter, see Alg.~\ref{alg:CKF}. It extracts key frames by filtering frames with slower hand motion and performing grouping deduplication operations. As shown in the lower left of Fig.~\ref{fig:framework}, with the input of hand ROI frames $X^{h}$ and hand position sequence $X^{p}$. 

Firstly, the filter evaluates hand motion speed by calculating the hand movement distance between adjacent frames:
\begin{equation}
 D_j = distance(X_{j-1}^{p},X_j^{p}), j\in[1,T),
\label{equa:distance}
\end{equation}
where $D_j$ indicates the hand moving distance between$X_{j-1}^{p}$ and $ X_j^{p}$, $distance(\cdot, \cdot)$ is the distance function, which is the Euclidean distance in this paper. 
Then, according to the moving distance of each frame and threshold $\sigma$, filter out the slow-motion frames:
\begin{equation}
 X_j^{h} \in S \mid D_j \leq \sigma,
\end{equation}
where $S$ is the slow-motion frame set. 

Secondly, to avoid repeated input of similar frames and meet the token length limit of MLLM, the deduplicate operation groups similar frames and takes the middle frame as the final keyframe. Slow-motion group formation based on the following two rules:

\textit{Rule 1.} Local Connection $R$: For $X_j^h, X_k^h \in S$, define $(X_j^h, R, X_k^h)$ if and only if $(|j - k| \le \theta)$. $\theta$ is the threshold for measuring frame index distance.

\textit{Rule 2.} Transitive Closure $R^*$: define $(X_j^h, R^*, X_k^h)$ if there exists a sequence of slow-motion frames in $S$ linking $X_j^h$ to $X_k^h$ with each consecutive pair satisfying Rule 1.

For any $(X_m^h \in S)$, the group is $G_m = {X^h\in S:(X^h,R^*,X^h_m)}$ or equivalently constructed via the recursive definition given above. After forming $M$ slow-motion groups, select the median frame of each group as the final $M$ keyframes and put into set $K$.

\noindent \textbf{Hand Position and Shape Support Set.}
Illustrate in mid-bottom of Fig.~\ref{fig:framework}, to exploit the in-context learning capability of MLLM, we construct a 40-frame support set from hand ROI frames in the training set of hearing cuers, which fully covers the 40 possible combinations of hand and position.

\noindent \textbf{Customized Prompt Template.}
In each prediction procedure, CCSPM put the filtered keyframes $K$ and support set into the customized prompt template. The prompt template is divided into four types of prompts according to its function. 

\textit{Background Prompts}. This prompt tries to let MLLM understand the role and background to be played. The template reveals that the MLLM should act as an expert in identifying the hand position and shape in the keyframes of the hand ROI in a CCS video. Then explained the background of the CS system and the role of hand coding in it.

\textit{Multimodal In-context Prompts}. With the help of the hand position and shape support set, this prompt attempts to use a multimodal approach of textual explanation plus image reference to fully utilize the multimodal in-context learning capability of MLLM~\cite{zeng2024can}. Specifically, the prompt first defines the five-category task of hand position and accurately describes the relative positions of the five categories (i.e., mouth, chin, throat, side, and cheek). The template also divides the support set into five categories according to the position label, so that each category has an 8-shot support set covering all hand shapes. The prompt defines the five categories in text and inputs support images in turn, in an attempt to establish a multimodal connection between the label, text, and visual information.

The prompt also uses a similar operation for the hand shape classification task. However, since the hand shape is not strictly defined in the system, we define eight different hand shapes based on the bending status of the five fingers. With the definition of different hand shapes, the corresponding 5-shot reference frames are also added to the final prompt.

\textit{Contrastive Prompts.} Since there are easily confused categories in both hand positions and shapes, to improve the recognition accuracy of MLLM, the template
compares the easily confused categories through text descriptions and provides methods to distinguish them. One example is \textit{``Shape 3 (Middle, ring, and pinky fingers are straight. Thumb and index fingers are bent.) is easily confused with label 4 (Index, middle, ring and pinky fingers are straight. Thumb is bent.) The difference between them is that in shape 3, the index finger is bent, and only three fingers are straight."}

\textit{Chain of thought Prompts.} 
For each keyframe in $K$, the template guides the MLLM to match hand positions/shapes with the support set via chain-of-thought reasoning, followed by label verification using textual definitions.
Following~\cite{wei2022chain}
The complete prompt template can be found in the Supplementary Materials.

\begin{table*}[t] \small
\caption{Details of different language CS Datasets. The ``Cuer'' row shows the different subset cuer settings in datasets, the format means the number of cuers and whether they are hearing-impaired. ``H'' means the cuers with normal hearing, and ``HI'' means it is a hearing-impaired dataset. Our newly proposed MHI-MCCSD fills the gap in large-scale multi-hearing-impaired cuer datasets.}
\centering

\tabcolsep 0.2in
\begin{tabular}{c|c|cc|ccc|c}
\toprule[1.3pt]
Dataset & French & \multicolumn{2}{|c|}{British} & \multicolumn{3}{|c|}{MCCSD} & MHI-MCCSD\\
\midrule[1pt]
Cuer & 1-H & 1-H & 5-H & 1-H & 1-HI & 6-H  & 8-HI\\
\midrule[1pt]
Sentence & 238 & 97 & 390 & 1000 & 818 & 6000 & 5272\\
Character &12872&2741&11021  &32902 & 8269 & 197412 & 152921\\
Word & -&- &- & 10564 & 8269 &63384 &49280\\
Phoneme &35&44&44&40&40&40&40 \\
Train&193&78&312&800&652&4800&4220\\
Test& 45&19&78&200&166&1200&1052\\
\bottomrule[1.3pt]
\end{tabular}

\label{table:comparative}
\end{table*}

\begin{table*}[t] \small
\caption{Comparison with other methods on MCCSD~\protect\cite{liu2023cross} and MHI-MCCSD. The table reports the CER and WER (\%) for different cuer settings. The ``Cuer'' row shows the different subset cuer settings in datasets, the format means the number of cuers and whether they are hearing-impaired. ``H'' means the cuers with normal hearing, and ``HI'' means it is a hearing-impaired dataset. Our STF-ACSR outperforms other methods by a large margin in all experimental settings.}
\centering

\tabcolsep 0.004 in
\begin{tabular}{c|cc|cc|cc|cc}

\toprule[1.3pt]
Cuer & \multicolumn{2}{|c|}{1-H}& \multicolumn{2}{|c|}{1-HI}& \multicolumn{2}{|c|}{6-H} &\multicolumn{2}{|c}{8-HI}\\
\midrule[1pt]
{\diagbox{Methods}{Metrics}} & CER $\downarrow$ & WER $\downarrow$ & CER $\downarrow$ & WER $\downarrow$ & CER $\downarrow$ & WER $\downarrow$ & CER $\downarrow$ & WER $\downarrow$ \\
\midrule[1pt]
ResNet18 +MHSA~\cite{vaswani2017attention} & 26.19 & 61.87 & 66.7 & 98.67 &61.83& 94.34& 82.03 & 99.91 \\
CMML~\cite{liu2023cross}& 9.81 & 25.54 & 32.23 & 69.45 & 30.01 &68.12  & 51.8& 91.26 \\
EcoCued~\cite{liu2024computation}& 9.54 & 25.03 &29.56 &61.59 & 29.75 & 67.83 &50.52 & 90.13\\

\midrule[1pt]
\textbf{STF-ACSR}& 
$\textbf{1.82}^{\textcolor{blue}{-80.9\%}}$&
$\textbf{5.19}^{\textcolor{blue}{-79.26\%}}$&
$\textbf{4.62}^{\textcolor{blue}{-84.37\%}}$&
$\textbf{12.21}^{\textcolor{blue}{-80.18\%}}$&
$\textbf{8.35}^{\textcolor{blue}{-71.93\%}}$&
$\textbf{21.06}^{\textcolor{blue}{-68.95\%}}$&
$\textbf{10.96}^{\textcolor{blue}{-78.3\%}}$&
$\textbf{25.67}^{\textcolor{blue}{-71.52\%}}$\\

\bottomrule[1.3pt]
\end{tabular}
\label{table:dataset}
\end{table*}

\subsection{Minimalist Fusion Module}
With the help of the CCSPM module, MLLM completes the recognition of the hand ROI keyframes and sends the results to MFM to complete the fusion with the lip-reading feature. MFM first embeds the recognition results, then makes the obtained matrix's dimension consistent with the lip feature dimension through a linear layer, and finally completes the fusion of hand and lip information by weighted addition.

\noindent \textbf{Embedding Function.} This function converts the hand recognition result into a matrix. Specifically, when the keyframe set containing M frames is recognized by MLLM, the predicted categories of position and shape are obtained for each frame. According to the Chinese CS system, each category corresponds to several phonemes. The embedding function takes the idea of one-shot encoding, and first initializes a matrix $H \in \mathbb{R}^{T \times 44} $ full of zeros, where $T$ is the frame number, 44 is the length of the word list which include 40 phonemes and four necessary symbols.
Second, for each keyframe $X^h_m$, according to the recognition result, the phoneme values corresponding to the vector $h_m \in \mathbb{R}^{1 \times 44}$ in the matrix are set to 1, and the same operation is taken for the frames belonging to the same group $G_m$ to ensure that the vectors corresponding to the frames in this group are all setting to 1 on the predicted phonemes.
Third, perform the same operation on all keyframes to complete the embedding of the hand movement recognition results and obtain the final hand embedding matrix $ H' \in \mathbb{R}^{T \times 44}$.

\noindent \textbf{Linear Layer and Fusion.} A single linear layer $f(\cdot)$ is used here to match the dimension of $H’$ with the dimension of the lip reading feature $L' \in  \mathbb{R}^{T \times d}$. These two features are then added together to form the final multimodal feature $P$:

\begin{equation}
\begin{aligned}
P = L'+ \lambda f(H'),
\end{aligned}
\end{equation}
where $\lambda$ is a learnable parameter.

\section{New Mandarin Chinese CS Dataset}
\label{sec:benchmark}

Previous work has found that hearing-impaired data have higher requirements for the ACSR system because of their different lip and hand movement habits from normal people. To further explore the effectiveness of the ACSR methods in real-world communication scenarios, we introduce the Multi-Hearing-Impaired Mandarin Chinese Cued Speech Dataset (MHI-MCCSD) as a supplement to the Mandarin Chinese Cued Speech Dataset (MCCSD)~\cite{liu2023cross}.  Following the construction of the MCCSD, we invited three female and five male Chinese hearing-impaired people to participate in the data recording. 
The recording environment is a silent room, The recording time spans several weeks, so for one cuer's data, it includes different lighting conditions and various appearance changes. The content contains 1010 different sentences, covering daily conversations, poems, idioms, and news broadcasts with lengths ranging from 4 to 25 words.

As shown in Tab.~\ref{table:dataset}, three cuers recorded the complete 1010 sentences, together with the other five cuers, MHI-MCCSD contains 5,272 sentences, which is the largest open-source multi-hearing-impaired dataset and provides the most comprehensive evaluation in a real hearing-impaired communication environment. Following MCCSD, MHI-MCCSD randomly split the data into a training set and a test set in a ratio of 4:1, with no overlapping sentences. Cued speech videos are in 1280 $\times$ 720 resolution and 30 frames per second.

Previous public CS benchmarks include MCCSD, French CS dataset~\cite{liu2018automatic}, and British English CS dataset~\cite{sankar2022multistream}. MCCSD contains three subsets, which are a six-normal-cuer setting containing 6000 sentences and two single-cuer settings recorded by a normal person and a hearing-impaired person, respectively. After MHI-MCCSD completed the supplement of multi-hearing-impaired data, MCCSD became the most comprehensive CS benchmark with its scale far exceeding other datasets. French CS dataset is a single-normal-cuer dataset, which contains 238 sentences. The French CS system has 35 different phonemes encoded by 8 hand shapes and 5 positions. British English CS dataset has both single and multi-cuer settings with 97 and 390 sentences, respectively. However, the multi-cuer setting is not open-sourced. In British English CS system, 44 phonemes are encoded by 8 shapes and 4 positions.

\begin{table*}[t] \small
\caption{Ablation study on hand information. The “Cuer” row shows the different subset cuer settings in datasets, the format means the number of cuers and whether
they are hearing-impaired. STF-ACSR effectively processed hand information to achieve significant performance improvement based on the lip-reading model.}
\centering
\begin{tabular}{c|cc|cc|cc|cc}
\toprule[1.3pt]
Cuer   & \multicolumn{2}{|c|}{1-H}& \multicolumn{2}{|c|}{1-HI}& \multicolumn{2}{|c|}{6-H} &\multicolumn{2}{|c}{8-HI} \\
\midrule[1pt]
{\diagbox{Methods}{Metrics}} & CER $\downarrow$ & WER $\downarrow$ & CER $\downarrow$ & WER $\downarrow$ & CER $\downarrow$ & WER $\downarrow$ & CER $\downarrow$ & WER $\downarrow$ \\
\midrule[1pt]
Pure Lip-reading &3.64& 10.36 &10.18 & 25.61&12.81 &32.74 &18.41 & 43.79\\
+Hand Infomation &1.82& 5.19 &4.62 &12.21& 8.35&21.06&10.96&25.67\\
\midrule[1pt]
Error reduction ratio &\textcolor{blue}{\textbf{-50$\%$}} &\textcolor{blue}{\textbf{-49.9$\%$}} &\textcolor{blue}{\textbf{-54.62$\%$}} &\textcolor{blue}{\textbf{-52.32$\%$}} & \textcolor{blue}{\textbf{-34.82$\%$}} &\textcolor{blue}{\textbf{-35.68$\%$}}& \textcolor{blue}{\textbf{-40.47$\%$}}& \textcolor{blue}{\textbf{-41.38$\%$}}\\

\bottomrule[1.3pt]
\end{tabular}
\label{table:ablation}
\end{table*}

\section{Experiments}

\label{sec:experiments}

\subsection{Experiment Setups}

\noindent \textbf{Datasets.} 
We conduct experiments on all MCCSD settings and the newly proposed MHI-MCCSD for a comprehensive evaluation of the effectiveness of STF-ACSR. The details of the dataset settings can refer to Tab.~\ref{table:dataset}.


\noindent \textbf{Implementations.}
For the data pre-processing, two open-source packages segment lip and hand ROIs from videos, i.e., retina-face~\cite{serengil2020lightface} and mediapipe2~\cite{lugaresi2019mediapipeframeworkbuildingperception}. For the lip-reading model, we use the pretrained VSR model from \cite{ma2023auto} with ResNet-18 front-end~\cite{he2016deep} and Conformer back-end~\cite{gulati2020conformer}. The dimension $d$ of the lip-reading feature is 768. For the hand-coding recognition, we use gpt-4o-2024-08-06~\cite{gpt4o} as the MLLM, which has powerful performance and full-featured APIs, to complete the task. Within the CCSPM, the slow-motion threshold $\sigma$ is set to 6, and the threshold 
for measuring frame index distance is set to 2. 
To avoid post-processing difficulties caused by messy output, we enabled the structured output function in the API, ensuring that GPT-4o predicts the hand and position categories of keyframes in integers and outputs them in JSON format.

In general, we use \cite{ma2023auto}'s settings to finetune the lip-reading model and the MFM's linear layer in a joint CTC/attention training manner~\cite{8068205}. All experiments are conducted on NVIDIA RTX A6000 GPUs.

\noindent \textbf{Evaluation Metrics.} We compare our STF-ACSR with other transformer-based ACSR methods, including CMML~\cite{liu2023cross} the first transformer-based method, and previous SOTA EcoCued~\cite{liu2024computation} which improves the computation and parameter efficiency. We also use Resnet18 + Multi-Head Self-Attention (MHSA)~\cite{vaswani2017attention} model as the baseline. The Character Error Rate (CER), calculated as the ratio of edit distance to total number of characters, and Word Error Rate (WER), similarly based on the word segmentation results, are used to evaluate performance from both phoneme and word aspects.

\begin{table}[t]
    \centering
    \caption{Ablation study on different prompts in MCCSD's customized prompt template and the support set. The table reports the classification accuracy (\%) of MLLM for hand position and shape in the hand keyframes on the MCCSD single-normal-cuer test set in different template settings. For best performance, we apply all prompts and the full support set in the customized prompt template.}
    \begin{tabular}{l|c c}
    \toprule[1.3pt]
    Prompts & Positions & Shapes\\
    
    \midrule[1pt]
         Background &55.41&51.18 \\
         + Multimodal In-context Prompts& 92.99 & 66.53\\
         + Contrastive Prompts &93.15 & 72.16 \\
         + Chain of Thought Prompts&94.87& 72.62\\
         + Positions and Shapes Support Set & 96.01 & 84.72 \\
    \bottomrule[1.3pt]
    \end{tabular}
    \label{tab:phases}
\end{table}

\subsection{Comparative Study}
The results of the comparison with other methods are shown in Tab.~\ref{table:comparative}. STF-ACSR achieves a significant performance lead in all experimental settings. In particular, it achieved an error rate reduction of more than 68\% compared to the second-place method, EcoCued, in all indicators. The amazing results fully demonstrate the effectiveness of STF-ACSR under the new paradigm design. Coming in second place across all experiments is the previous SOTA, EcoCued, but only slightly ahead of CMML.

It is particularly noteworthy that previous methods showed a significant performance degradation in hearing-impaired experiments compared with the hearing experiments, while STF-ACSR maintained similar performance. This proves that STF-ACSR achieves robust performance in real hearing-impaired environments by integrating a large-scale pre-trained lip reading model and using MLLM with strong generalization ability for hand-coding recognition.

\subsection{Ablation Study}
\begin{figure*}[t]
\includegraphics[width=0.8\linewidth]{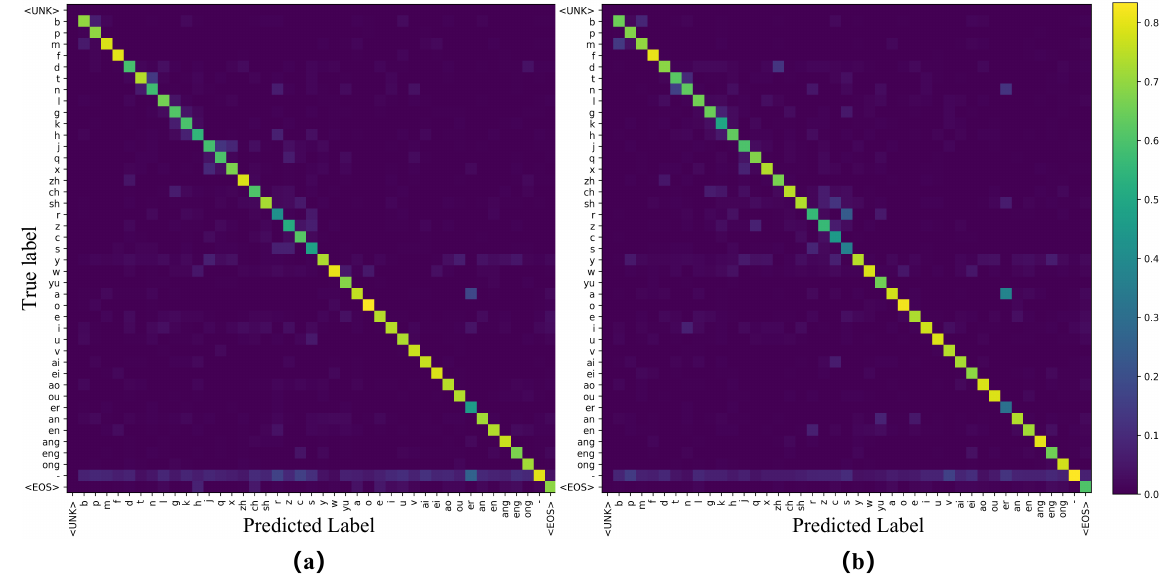}
\centering
\caption{Phoneme confusion matrices of STF-ACSR model on the test sets of multi-normal-cuer setting of MCCSD~\protect\cite{liu2023cross} (six-hearing cuers, shows in subfigure.~(a)) and MHI-MCCSD (eight hearing-impaired cuers, shows in subfigure.~(b)). Results reveal phonemes that are easily confused.}
\label{fig:confusion} 
\vspace{-0.8em}
\end{figure*}

\noindent \textbf{Effect of Hand Information.} In STF-ACSR, lip ROI frames are processed by a pretrained lip-reading model and the hand ROI frames are processed using the train-free recognition capability of MLLM with the help of CCSPM. We remove the hand information from MCCSD to evaluate its effectiveness. Results shows in Tab.~\ref{table:ablation}, based on the performance of pure lip reading, hand information reduces the error rate by more than 30\% in all indicators. This shows the effectiveness of STF-ACSR, which simplifies the complex hand recognition task and the fusion module that requires a special design into a hand keyframe recognition task through CCSPM and uses MFM for effective information fusion.

It is worth noting that in the hearing-impaired experimental settings, hand information brings more performance gains. This phenomenon is consistent with the fact that it is challenging to recognize hearing-impaired people‘s lip movements and fully demonstrates the effectiveness of STF-ACSR in processing hand information.

\noindent \textbf{Effect of Prompts in Prompt Template.}
In Tab.~\ref{tab:phases}, we compare the impact of different prompts in the customized prompt template. We use the single-normal-cuer (i.e., 1-H) test set of MCCSD to evaluate MLLM‘s classification accuracy of the hand position and shape of the hand keyframe under different template settings. 

Specifically, we used the classification results of keyframes with the frame labels in the dataset for verification. Using template just involving the background prompts can only achieve 55.41\% and 51.18\% accuracy in position and shape classification, respectively. With the addition of the multimodal in-context prompts, the accuracy increases to 92.99\% and 66.53\% respectively. It is worth noting that the multimodal in-context prompts here do not include the complete positions and shapes support set, but only use a single frame as a reference while defining the position and shape labels in text. With the addition of the Contrastive prompts and the Chain of Thought prompts the prediction accuracy of MLLM gradually improved. Finally, with the help of the complete positions and shapes support set, the classification accuracy of hand position and shape increased to 96.01\% and 84.72\%, respectively. The results demonstrate the effectiveness of prompts within the customized prompt template in CCSPM and the necessity of a complete support set.

\noindent \textbf{Discussion of Hearing and Hearing-Impaired. }
In order to intuitively analyze the impact of normal hearing data and hearing-impaired scenarios on the ACSR model, we visualized the results of STF-ACSR on the MCCSD 6-hearing cuers and MHI-MCCSD test sets and displayed them in the form of phoneme confusion matrices shown in Fig.~\ref{fig:confusion}. The coordinates of the matrix include 40 Chinese CS phonemes and necessary symbols. The vertical and horizontal axes of the matrix represent true phonemes and model predictions, respectively. By observing the obvious color blocks that deviate from the diagonal in the matrix, we can get the phonemes that are easily confused by the model in prediction. 

The results show models are most likely to mispredict \textit{`a’} as \textit{`er’} in both normal cuer and hearing-impaired cuer environments, and this phenomenon is more obvious in hearing-impaired scenarios. According to the Chinese CS system, these two phonemes are represented by the same hand shape in hand coding, so this error comes from the easily confused mouth shape when reading these two vowels.

In addition, in the hearing-impaired environment, the model tends to mispredict \textit{`r’} as \textit{`s’} because it can only distinguish them from lip reading. However, there are also confusing pairs caused by similar hand and lip movements together, such as \textit{`b’} and \textit{`m’}, whose corresponding hand shapes can only be distinguished by whether the thumb is bent.

\section{Conclusion}
\label{sec:conclusion}
The proposed STF-ACSR introduces a novel semi training-free paradigm for ACSR, effectively addressing the challenges of limited data and overfitting in hand feature extraction. By leveraging the zero-shot capabilities of MLLMs through the CCSPM, the method simplifies hand movement recognition into keyframe-based classification and integrates it with lip-reading via MFM. The new MHI-MCCSD dataset, which includes data from eight hearing-impaired cuers, further enhances the robustness and practicality of ACSR systems. Extensive experiments demonstrate that STF-ACSR achieves SOTA performance across both normal and hearing-impaired scenarios, with significant error rate reductions (up to 84.37\% CER improvement) compared to existing methods. This work not only advances ACSR technology but also promotes barrier-free communication for the hearing-impaired community by bridging the gap between visual cues and spoken language understanding. Furthermore, under the rules of existing CS systems, STF-ACSR demonstrates its potential for simple and effective generalization to other language-specific CS systems, which will be realized in future work.
\clearpage

\bibliographystyle{named}
\bibliography{ijcai25}

\end{document}